\documentclass[12pt,preprint]{aastex}
\usepackage{apjfonts}
\usepackage{natbib}
\usepackage{amssymb, amsmath, amsbsy, epsfig, epsf}

\newcommand{\lum}{ergs s\ensuremath{^{-1}}}
\newcommand{\lbol}{\ensuremath{L\mathrm{_{bol}}}}

\newcommand{\ledd}{\ensuremath{L\mathrm{_{Edd}}}}
\newcommand{\lratio}{\ensuremath{L/\ledd}}

\newcommand{\msun}{\ensuremath{M_{\odot}}}
\newcommand{\kms}{\ensuremath{\mathrm{km~s^{-1}}}}
\newcommand{\mbh}{\ensuremath{M_\mathrm{BH}}}

\newcommand{\rs}{\ensuremath{r_{\rm \scriptscriptstyle S}}}
\newcommand{\pnull}{ \ensuremath{P_{\mathrm{null}}} }

\newcommand{\hb}{H\ensuremath{\beta}}

\newcommand{\hi}{H\,{\footnotesize I}}

\newcommand{\feii}{Fe\,{\footnotesize II}}

\newcommand{\mgii}{Mg\,{\footnotesize II}}

\newcommand{\civ}{C\,{\footnotesize IV}}
\newcommand{\nv}{N\,{\footnotesize V}}

\newcommand{\colnh}{\ensuremath{N_\mathrm{H}}}

\slugcomment{To Appear in {\it The Astrophysical Journal Letters}}
\shorttitle{\lratio\ governs \mgii\ equivalent width}
\shortauthors{Dong et al.}

\begin{document}

\title{Eddington ratio governs the equivalent width of \mgii\ emission line
in active galactic nuclei}

\author{Xiao-Bo Dong\altaffilmark{1,4},
Ting-Gui Wang\altaffilmark{1,4}, Jian-Guo Wang\altaffilmark{2,1,7},
Xiaohui Fan\altaffilmark{3,5},\\ Huiyuan Wang\altaffilmark{1,4},
Hongyan Zhou\altaffilmark{1,4,6}, and Weimin Yuan\altaffilmark{2} }

\altaffiltext{1}{Key Laboratory for Research in Galaxies and Cosmology,
The University of Sciences and Technology of China, Chinese Academy of Sciences,
Hefei, Anhui 230026, China; ~ xbdong@ustc.edu.cn}
\altaffiltext{2}{National Astronomical Observatories/Yunnan
Observatory, Chinese Academy of Sciences, P.O. Box 110, Kunming,
Yunnan 650011, China; ~ wmy@ynao.ac.cn}
\altaffiltext{3}{Steward Observatory, The University of Arizona, Tucson,
AZ 85721, USA; ~ fan@as.arizona.edu}
\altaffiltext{4}{Center for Astrophysics, University of Science and
Technology of China, Hefei, Anhui 230026, China;}
\altaffiltext{5}{Max Planck Institute for Astronomy, D-69120, Heidelberg, Germany}
\altaffiltext{6}{Max-Planck-Institut f\"ur extraterrestrische
Physik, Giessenbachstrasse 1, 85748 Garching, Germany}
\altaffiltext{7}{Graduate School of the Chinese Academy of Sciences,
19A Yuquan Road, P.O.Box 3908, Beijing 100039, China}

\begin{abstract}
We have investigated the ensemble regularities of the equivalent widths (EWs) of
\mgii\ $\lambda 2800$ emission line of active galactic nuclei (AGNs),
using a uniformly selected sample of 2092 Seyfert 1 galaxies and quasars
at $0.45 \leq z \leq 0.8$
in the spectroscopic data set of
Sloan Digital Sky Survey Fourth Data Release.
We find a strong correlation between the EW of \mgii\ and the AGN Eddington ratio
(\lratio): ${\rm EW(MgII)} \propto (\lratio)^{-0.4}$.
Furthermore, for AGNs with the same \lratio,
their EWs of \mgii\ show no correlation with
luminosity, black hole mass or line width,
and the \mgii\ line luminosity is proportional to continuum
luminosity, as expected by photoionization theory.
Our result shows that
\mgii\ EW is not dependent on luminosity, but is solely governed by \lratio.
\end{abstract}


\setcounter{footnote}{0}
\setcounter{section}{0}

\section{Introduction}

\mgii\ $\lambda$2800 is one of the most prominent broad emission lines
in the near-ultraviolet spectra of type-1 active galactic nuclei
(AGNs, including Seyfert galaxies and quasars).
Its equivalent width (EW) is generally in the range of about 10--100\AA\
(Dietrich et al. 2002).
Theory shows that
as a low-ionization line,
\mgii\ originates
from optically thick (i.e. large column density) clouds only
that suffer little from radial motions,
thus it is valuable in probing the properties of the clouds gravitationally
bound in the AGN broad-line region (BLR),
and in estimating the  virial mass of the central supermassive black holes
(e.g. Grandi \& Phillips 1979, McLure \& Dunlop 2004,
Wang et al. 2009).

However, \mgii\ is highly blended with the \feii\ multiplet emission.
This complicates accurate \mgii\ measurement in AGN spectra, and makes
a comprehensive study  difficult of \mgii\ using
large, homogeneous AGN samples.
By composite-spectrum analysis,
Dietrich et al. (2002) found a negative correlation
between the \mgii\ EW and the continuum luminosity,
similar to the Baldwin (1977) effect first discovered in \civ\ line.
However, except that and some others using small samples (e.g., Grandi \& Phillips 1979,
Zheng et al. 1992, Espey \& Andreadis 1999, Croom et al. 2002),
there has been few studies so far on the \mgii\ properties using
a large, homogeneous AGN sample.

Recently, this technical difficulty has been overcome by
the availability of the detailed \feii\ templates in the wavelength region
covered by \mgii\
(e.g., Tsuzuki et al. 2006; cf. Vestergaard \& Wilkes 2001).
In this Letter, we explore the regularities of the \mgii\ EW
in the AGN ensemble,
by taking advantage of the unprecedented spectroscopic data
from the Sloan Digital Sky Survey (SDSS; York et al. 2000) and the new \feii\ templates.
We find that, at the 0th-order,
\mgii\ luminosity is directly proportional to continuum luminosity;
at the 1st-order,
the proportional coefficient correlates negatively with the Eddington ratio
($\ell \equiv \lratio$).%
\footnote{Eddington ratio ($\ell \equiv \lratio$) is
the ratio between the bolometric and Eddington luminosities.
Eddington luminosity (\ledd), by definition, is the luminosity at which the gravity
of the central source acting on an electron--proton pair (i.e. fully ionized gas)
is balanced by the radiation pressure due to electron Thomson scattering.
Thus \emph{Eddington ratio} is different from \emph{relative/normalized accretion rate}
both in meaning and in scope of application;
see Dong et al. (2009) for a detailed discussion.
For clarification and for the ease of notation,
below we use \lratio\ or $\ell$ alternately.}
Throughout this paper, we use a cosmology with
$H_{\rm 0}$=70 km\,s$^{-1}$\,Mpc$^{-1}$, $\Omega_{\rm M}$=0.3 and
$\Omega_{\rm \Lambda}$=0.7.

\section{Sample and Data analysis}

We use the UV sample of 2092 Seyfert 1 galaxies and quasars of Dong et al. (2009),
selected from the spectral data set of the SDSS Fourth Data Release
(Adelman-McCarthy et al. 2006).
Sample definition and data analysis methods are described in detail in that work.
Briefly, we select quasars with both
the \hb\ and \mgii\ present in the SDSS spectra
and with  continuum and emission lines suffering little
from the contamination of the host-galaxy starlight.
The criteria are:
(a) $0.45 \leq z \leq 0.8$, (b) the median signal-to-noise ratio (S/N) $\geq 10$ per pixel,
and (c) the rest-frame absorption-line EWs of Ca\,K (3934\AA),
Ca\,H + H$\epsilon$ (3970\AA) and H$\delta$ (4102\AA) $< 2\sigma$.
The \hb\ BLR has been extensively studied using reverberation mapping for about forty type-1 AGNs,
and the black hole mass formalisms based on single-epoch \hb\ are
well calibrated with reverberation mapping data
(Peterson et al. 2004, Bentz et al. 2006, Bentz et al. 2009).
Thus the presence of \hb\ in our sample is helpful to double check
the findings in this Letter (see \S4).

As described in Dong et al. (2009),
 the SDSS spectra are  corrected for Galactic extinction
and the \mgii\ emission is  separated from AGN continuum,
Balmer continuum and the \feii\ multiplets.
Then the broad components of the \mgii\,$\lambda\lambda2796,\,2803$ doublet lines
are each modeled with a truncated 5-parameter Gauss-Hermite series profile
(see also Salvainder et al. 2007).
The narrow component of each line is fitted with a Gaussian.
In the fitting, the narrow component is constrained as follows:
FWHM $\leq 900$ \kms\ and flux $< 10$ per cent of the total \mgii\ flux
(Wang et al. 2009; cf. Wills et al. 1993).
UV \feii\ is modeled with the tabulated semiempirical template
generated by Tsuzuki et al. (2006) based on their measurements of I\,ZW\,1;
in the wavelength region covered by the \mgii\ emission,
this template uses the calculation with the CLOUDY
photoionization code (Ferland et al. 1998).

All the data used in this Letter are presented in Dong et al. (2009).
The detailed fitting results are available online
for the decomposed spectral components
(continuum, \feii\ multiplets and other emission lines).%
\footnote{Available at
http://staff.ustc.edu.cn/\~{}xbdong/Data\_Release/ell\_effect/,
together with auxiliary code to explain and demonstrate the fitting
and the parameters.}
We compared the \mgii\ parameter values fitted by our fitting method
with those by the methods of McLure \& Dunlop (2004) and Salviander et al. (2007).
Our results are well consistent with those obtained by
Salviander et al., and roughly consistent with those obtained by
McLure \& Dunlop  (Wang et al. 2009;
cf. Salviander et al. 2007).

We estimate the measurement uncertainties of the parameters
using a bootstrap method
(Wang et al. 2009; cf. Dong et al. 2008).
The estimated typical 1$\sigma$ relative errors
are 10\% and 8\%, respectively, for the fluxes of broad \mgii\ and \hb;
20\% and 15\%, respectively, for the FWHM of broad \mgii\ and \hb;
8\% and 5\%, respectively, for the slope and normalization of the local continua.
According to standard propagation of errors,
the formal 1$\sigma$ error of \mgii\ EW (in log-scale) is typically 0.05 dex.
We must note that the thus estimated errors still do not account for the uncertainties
caused by \feii\ template mismatch, etc., and that it is almost impossible to pin down
the real measurement errors for individual objects.
Hence we waive the analysis on the intrinsic scatter of
the relations of interest in this Letter.

We calculate the black hole masses, \mbh, based on the \mgii\ FWHM and
the monochromatic luminosity $L_{3000} \equiv \lambda L_{\lambda}$(3000\AA)
using the formalism of McLure \& Dunlop (2004).
The small discrepancy of the fitted \mgii\ parameters between our method and
the method of McLure \& Dunlop (2004) has no significant effect on our conclusions
(see also \S4).
The Eddington ratios are calculated assuming
a bolometric luminosity correction $\lbol \approx 5.9 L_{3000}$
(McLure \& Dunlop 2004).
The mean and standard deviation (computed in log-space) of
\mgii\ FWHM are 3200 \kms\ and 0.18 dex;
$L_{3000}$, $1.2 \times 10^{45}$\lum\ and 0.31 dex;
\mbh, $1.6 \times 10^{8}$\msun\ and 0.42 dex;
\lratio, 0.39 and 0.38 dex.

\section{Results}
\subsection{Correlations of \mgii\ EW with other quantities}
We use the Spearman rank method
to perform bivariate correlation analysis of \mgii\ EW with
\mgii\ FWHM, $L_{3000}$, \mbh\ and \lratio.
The results are summarized in Table 1 (top panel),
giving the Spearman coefficient (\rs) and the two-tailed probability
($P_{\rm null}$) that a correlation is not present.
The strongest correlation of \mgii\ EW is with \lratio\ ($\rs = 0.59$)
which is a combination of \mgii\ FWHM and $L_{3000}$
as $\lratio \propto {\rm FWHM}^{-2} L_{3000}^{0.5}$.
The correlation of \mgii\ EW with another combination, \mbh\
($\propto {\rm FWHM}^{2} L_{3000}^{0.5}$), is much weaker ($\rs = 0.38$).
The correlation with FWHM ($\rs = 0.55$) is slightly weaker than
that with \lratio, while
the correlation with $L_{3000}$ is weak ($\rs = 0.2$).
These correlations are illustrated in Fig. 1.

We must note that because the SDSS spectroscopical survey is magnitude-limited,
\mgii\ FWHM, $L_{3000}$, \mbh\ and \lratio\ in our sample
correlate one with another apparently.
The apparent (likely not intrinsic) correlation between \mbh\ and \lratio\
is further enhanced by the correlation of their measurement errors,
because both \mbh\ and \lratio\ are constructed from
\mgii\ FWHM and $L_{3000}$.
The Spearman correlation coefficients (\rs) of \lratio\
with \mgii\ FWHM, $L_{3000}$ and \mbh\ are $-0.92$, $0.27$ and $-0.64$, respectively.

In light of the serious inter-dependence among these four quantities,
the correlations of \mgii\ EW with the first three quantities
is probably only the secondary effect of the strongest (thus presumably intrinsic)
correlation with \lratio.
To test this possibility, we perform partial Spearman rank correlation analysis
(Kendall \& Stuart 1979). The results are summarized in Table 1.
When controlling for \lratio, none of the correlations of \mgii\ EW with
FWHM, $L_{3000}$ or \mbh\ is significant at $\pnull < 0.01$;
yet controlling for these three quantities, the correlation of \mgii\ EW with \lratio\
is still significant with $\pnull < 10^{-38}$.
This fact suggests that \lratio\ is
the \emph{sole} parameter, at least among the four, that governs the EW of \mgii.

From Fig. 1 there appears a linear relation between \mgii\ EW and \lratio\
in log--log scale. We perform linear regression,
treating \lratio\ as the independent variable, and  find:
\begin{equation}
 \log {\rm EW(\mgii)}  ~=~
  ( 1.26 \pm 0.02)   ~+~  (-0.39 \pm 0.03) \log \ell ~~.
\end{equation}
The errors on the fit parameters are purely statistical, and the actual uncertainty
could be somewhat larger (cf. \S2).

\subsection{The dependence of \mgii\ EW:
not on luminosity, but on Eddington ratio only}

As suggested by the above partial correlation tests,
the \mgii\ EW likely does \emph{not} depend on the continuum luminosity or
black hole mass, but on \lratio\ only as in Eqn. 1.
That is, (1) the emission-line luminosity, $L_{\rm \mgii}$,
is directly proportional to the continuum luminosity at a fixed \lratio,
while (2) the proportional coefficient
is different for objects with different \lratio.

To further investigate the \mgii\ EW--$L$ relation,
we take an alternative approach to analyze this problem.
In Fig. 2 we plot the distribution
of $L_{\rm \mgii}$ versus $L_{3000}$ using the objects in our sample,
with the 433 objects of $\lratio \geq 0.8$ (high-$\ell$ subsample) denoted in blue
and the 456 objects of $\lratio \leq 0.2$ (low-$\ell$ subsample) in red.
It is evident that the objects of the high-$\ell$ subsample are systematically
smaller than those of the low-$\ell$ subsample in $L_{\rm \mgii}$ at
any given $L_{3000}$.
We perform linear regression in log scale
with $L_{3000}$ being taken as the independent variable,
according to
\begin{equation}
 \log L_{\rm \mgii}  ~=~
  k \, \log L_{3000}   ~+~  b ~~.
\end{equation}
We fit the relation for the full sample, the high-$\ell$ and low-$\ell$ subsamples,
and the intermediate-$\ell$ subsample with $0.2 < \lratio < 0.8$, respectively,
with and without $k$ being fixed to be 1. The results are summarized in Table 2.
The low-$\ell$ subsample has 456 objects clustering mostly
in a narrow range of luminosity,
yet with a fairly large dynamical range in \lratio\ of about 1 order of magnitude.
The fit with $k$ being an additional free parameter for this subsample
is not statistically guaranteed according to $F$-test.
Thus in Table 2 we list only the fitting result with $k$ fixed to be 1.
Instead, we select the subsample of 299 objects with $0.1 < \lratio \leq 0.2$
and perform the fitting. The results are listed in Table 2 too.

The best-fit $k$ values for the high-$\ell$ and intermediate-$\ell$ subsamples
are well consistent with 1 within 1$\sigma$.
For the $0.1 < \lratio \leq 0.2$ subsample,
the best-fit $k$ is consistent with 1 within 1.5$\sigma$.
As \mgii\ EW is equivalent to $L_{\rm \mgii}/L_{3000}$,
such a direct proportional relation between  $L_{\rm \mgii}$ and $L_{3000}$
means that the \mgii\ EW does not depend on luminosity.

The mean values of $\log \lratio$ of the full sample,
the high-$\ell$, intermediate-$\ell$,
$0.1 < \ell <0.2$ and low-$\ell$ subsamples
are $-0.41$, 0.08, $-0.38$, $-0.83$ and $-0.95$, respectively.
The coefficients ($b$) of these best-fit proportion ($k=1$) relations
are consistent with the predictions of Eqn. 1.

The best-fit $k$ ($0.91\pm0.01$) for the full sample is significantly lower than 1,
which means an apparent dependence of \mgii\ EW on luminosity for the full sample,
as illustrated in Fig.1.
However, this can be naturally explained by the selection effect mentioned in \S3.1:
high-$\ell$ (and thus small-EW) objects dominate the high-luminosity end while
low-$\ell$ (and thus large-EW) objects tend to cluster in
the low-luminosity end, which is clearly displayed in Fig. 2.
As a demonstration (see also \S4), in the panel b of Fig. 1
we over-plot the inferred \mgii\ EW--$L_{3000}$ relation,
the Baldwin effect, that
corresponds to the fitted $\log L_{\rm \mgii} - k \log L_{3000}$
relation where $k=0.91$.

\section{Discussions and conclusions}

To check possible effects of  black hole estimation on
the above correlations,
we re-examine the above correlation tests by calculating \mbh\
using all the other available formalisms based on broad \hb\ and \mgii\ lines
(see a compilation of the formalisms in McGill et al. [2008]
and the formalisms in Wang et al. [2009] and Vestergaard \& Osmer [2009]).
All the tests give similar results to those listed in Table 1 (top panel).
This is mainly because the \mbh\ dynamical range covered by our sample
is not large ($\approx 1.5$dex), and the various formalisms based on single-epoch
\hb\ or \mgii\ have only subtle differences one from another
(Wang et al. 2009).
In the bottom panel of Table 1, we also list the results
using the FWHM of broad \hb, the luminosity $\lambda L_{\lambda}$(5100\AA),
the black hole masses calculated with the formalism presented in
Collin et al. (2006, their Eqn. 7)
and the Eddington ratios assuming $\lbol \approx 9\lambda L_{\lambda}$(5100\AA).

Considering the fairly narrow dynamical range of \lratio\
($1\sigma = 0.38$ dex in log-scale) and the measurement errors,
such a significant correlation between \mgii\ EW and \lratio\ is surprising,
suggesting a rather tight intrinsic relationship between them.
More importantly, \mgii\ EW is likely to be intrinsically solely dependent
on \lratio\, but \emph{not} on continuum luminosity.
The dynamical range of the \mgii\ EW of the objects in our sample is
similar to that of the composite spectra of Dietrich et al. (2002).
They reported a  Baldwin effect of \mgii\ as
$\log {\rm EW} \propto (-0.09\pm0.01) \log L$.
This slope is exactly consistent with our $k$ value (0.91;
see Table 2 and Fig. 1b)
for the full sample, which is likely due to the selection effect discussed above.
However, the range of our $L_{3000}$
is $2\times 10^{44} - 2\times 10^{46}$ \lum\
whereas that of Dietrich et al. (2002)
is $2\times 10^{42} - 2\times 10^{47}$ \lum\
(converted from $\lambda L_{\lambda}$(1450\AA) by assuming
$f_{\lambda} \propto \lambda^{-1.5}$).
Our sample lacks objects of relatively low luminosity.
Thus, if our above finding holds for the AGN ensemble
in the full luminosity range as in Dietrich et al. (2002),
then the classical Baldwin effect of \mgii\ is purely a secondary effect of the
EW--$\ell$ relation in combination with
the selection effect inherent in any flux-limited sample
(see the analysis in the last paragraph of \S3.2).

It is intriguing to understand the EW--$\ell$ relation
as expressed in Eqn. 1 (and Table 2;
cf. Baskin \& Laor 2004, Bachev et al. 2008).
First of all, the fact that the \mgii\ luminosity is directly proportional to
continuum luminosity is just as expected from the  photoionization theory.
Second, in the photoionization picture (Osterbrock \& Ferland 2006),
the emission-line EW is dependent mainly on two parameters,
the shape of the AGN continuum and
the covering factor of the line-emitting clouds,
thus the fact that the proportional coefficient
is different for different \lratio\
suggests that either the continuum shape or the covering factor or both
is dependent on \lratio.

Zheng \& Malkan (1993) and others have proposed a model in which
systematic variation in the ionizing continuum shape with luminosity explains
the Baldwin Effect, via the softening of the accretion disk spectrum with
increasing black hole mass.
More recently, Korista et al. (1998) generated grids of photoionization
models to demonstrate this relationship quantitatively, and
additionally invoked variations in gas metallicity with luminosity to
explain the peculiar weakness of the Baldwin Effect in \nv~$\lambda$1240.
The phenomenological model of Korista et al.\ relating ionizing continuum shape and
gas metallicity with quasar luminosity (or \mbh) generally explains
the Baldwin Effect in
almost all emission lines (cf.\ Dietrich et al.\ 2002).
The remaining question is the underlying physical link
between the shape of the ionizing continuum and the fundamental parameters
of the accretion process such as \mbh\ and normalized accretion rate
(i.e. \lratio, as the accretion rate is not an observable), which
is still not clear (see, e.g., \S6 of Dietrich et al. 2002).
There are observational reports that the ratio of the X-ray to
the bolometric/optical luminosity
correlates negatively with \lratio\
(Vasudevan \& Fabian 2007, Kelly et al. 2008).
This might account for the \mgii\ EW--$\ell$ relation as follows:
as \lratio\ increases, the ratio of the X-ray to UV photons decreases, i.e.,
the heating of the \hi$^{*}$ region weakens and thus
the \mgii\ EW decreases. We will use other emission lines such as \feii\
to explore this issue in a following paper (Dong et al. 2009).

Alternatively, we propose a physical model
involving cloud properties (effectively the covering factor).
The details of the model are to be presented in the following paper (Dong et al. 2009);
here we only present a brief description.
As described in \S1, \mgii\ line originates only from a thin
transition layer in the partially ionized \hi$^{*}$ region
of the ionization-bounded (high-\colnh) clouds
(see Collin-Souffrin et al. [1986] for the details).
On the other hand, because low-\colnh\ clouds are not massive enough to balance
the radiation pressure force that
is boosted due to photoelectric absorption
as large as by about one or more orders of magnitude,
they are blown out of the AGN BLR even at small \lratio.
(see Marconi et al. 2008, 2009; also Fabian et al. 2006, but cf. Netzer 2009).
According to the photoionization calculation
of Fabian et al. (2006, see their Fig. 1),
for dust-free clouds of $\colnh \gtrsim 10^{21}$ cm$^{-2}$ with
photoionization parameter $U \lesssim 1$,
the lower limit of the column density of the clouds that
can survive in the BLR is approximately proportional to \lratio,
as $\colnh > 10^{23} \, \ell$ cm$^{-2}$;
for dusty clouds, $\colnh > 5 \times 10^{23} \, \ell$ cm$^{-2}$.
That is, the higher the \lratio\ of an AGN is,
the larger is the fraction of high-\colnh\ clouds that are accumulated in the BLR.
So, for \mgii\ line, it is the ionization-bounded clouds
\emph{gravitationally bound in the BLR}
that emit it predominantly.
Thus, in the AGN ensemble, as \lratio\ increases, the number of the clouds that emit
\mgii\ efficiently, in other words, the effective covering factor,
decreases, and so does the \mgii\ EW.
Thus the model explains the EW--$\ell$ relation naturally.
The distinguishing of the two models requires detailed photoionization modeling,
which is beyond the scope of this Letter.

\acknowledgments
We thank the anonymous referee for his/her helpful comments and suggestions.
DXB thanks Kirk Korista for his careful reading and
valuable suggestions that improved this paper,
and thanks Alessandro Marconi and Luis Ho
for the helpful discussions and comments.
This work is supported by Chinese NSF grants
NSF-10533050, NSF-10703006 and NSF-10728307,
a CAS Knowledge Innovation Program (Grant No. KJCX2-YW-T05),
and a National 973 Project of China (2007CB815403).
Funding for the SDSS and SDSS-II has been provided by the Alfred P. Sloan Foundation,
the Participating Institutions, the National Science Foundation,
the U.S. Department of Energy, the National Aeronautics and Space Administration,
the Japanese Monbukagakusho, the Max Planck Society,
and the Higher Education Funding Council for England.
The SDSS Web Site is http://www.sdss.org/.



\setcounter{figure}{0}
\setcounter{table}{0}

\figurenum{1}
\begin{figure}[tbp]
\epsscale{1.} \plotone{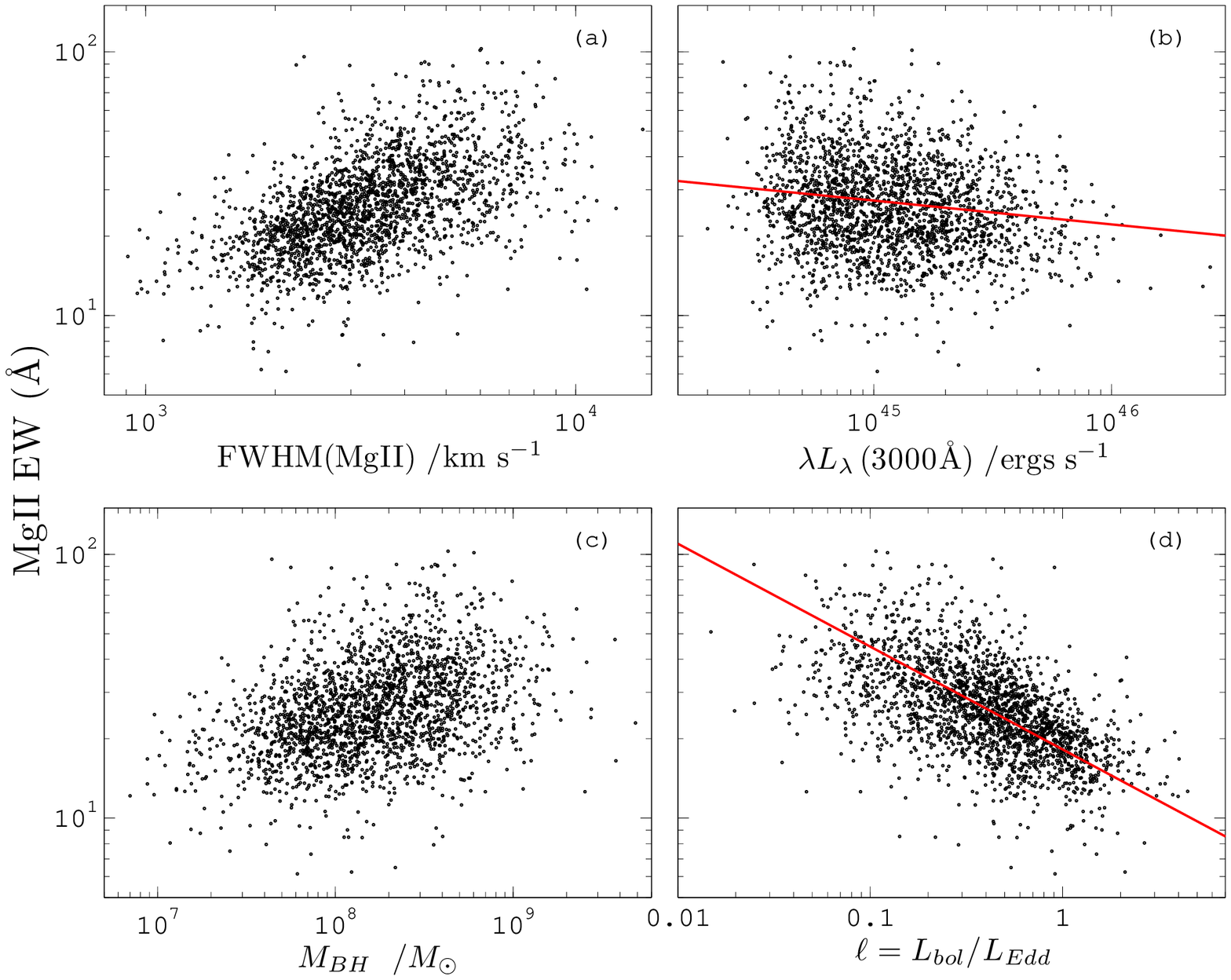} \caption{ Plots of the equivalent
width of \mgii\ $\lambda 2800$ emission line of the 2092 type-1 AGNs
in our sample versus their \mgii\ FWHM, $\lambda
L_{\lambda}$(3000\AA), black hole mass (\mbh), and Eddington ratio
($\ell \equiv \lbol/\ledd$). In panel b we also plot the apparent
relation of the \mgii\ Baldwin effect converted from the best-fit
$\log L_{\rm \mgii} - \log L_{3000}$ relation with a slope of 0.91
(see Table 2), which is consistent with that of Dietrich et al.
(2002). In panel d we also plot the best-fit linear relation in
log--log scale between \mgii\ equivalent width and \lratio\ (Eqn.
1). }
\end{figure}

\figurenum{2}
\begin{figure}[tbp]
\epsscale{1.} \plotone{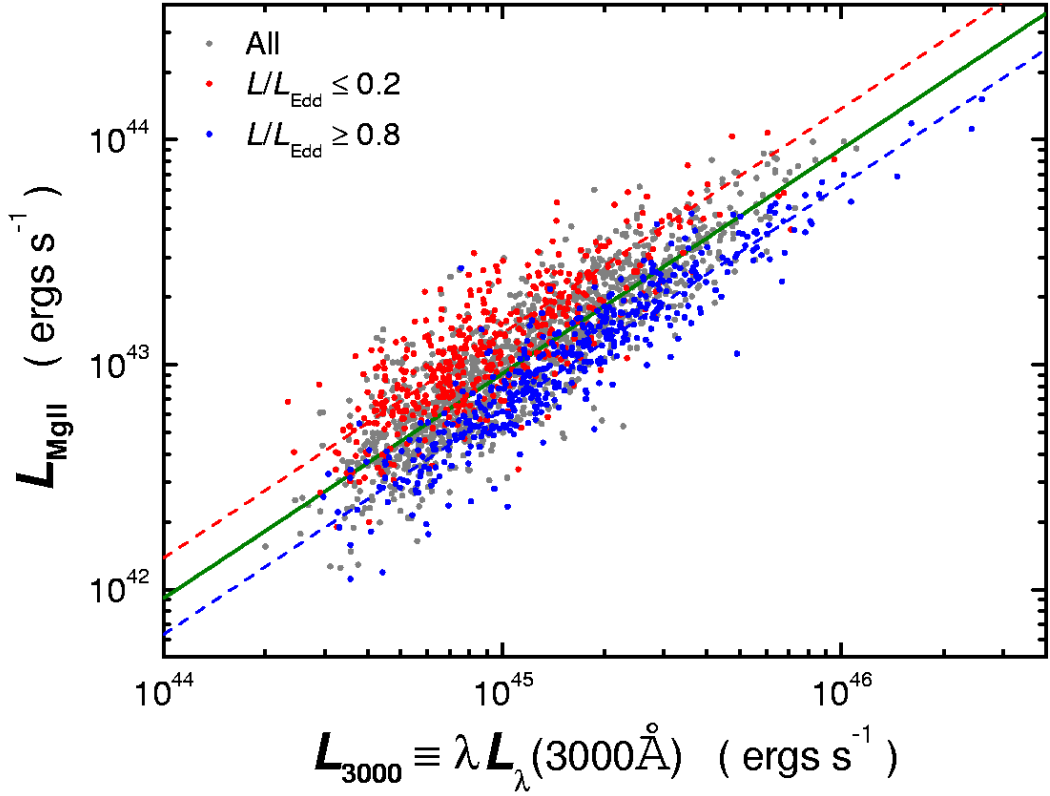} \caption{ \mgii\ luminosity versus
continuum luminosity $\lambda L_{\lambda}$(3000\AA) of the type-1
AGNs in our sample. Objects with $\lratio \geq 0.8$ are denoted in
blue and those with $\lratio \leq 0.2$ in red. Also plotted are the
best-fit proportion relations for the full sample (green line), the
subsample of $\lratio \geq 0.8$ (blue line) and the subsample of
$\lratio \leq 0.2$ (red line), see \S3.2 and Table 2.}
\end{figure}

\begin{deluxetable}{lllllllllll}
\tablenum{1} \tablewidth{0pt} \topmargin 0.0cm \evensidemargin = 0mm
\oddsidemargin = 0mm
\tabletypesize{\scriptsize}
\tablecaption{Results of Spearman correlations and partial correlations of \mgii\ EW
   \tablenotemark{~a} }
\tablehead{                  & \colhead{FWHM} &
           \colhead{$L$} &
           \colhead{\mbh} &
           \colhead{\lratio} &
           \colhead{($\ell$;~FWHM) \tablenotemark{~b}} &
           \colhead{($\ell$;~$L_{3000}$) \tablenotemark{~b}} &
           \colhead{($\ell$;~\mbh) \tablenotemark{~b}} &
           \colhead{(FWHM;~$\ell$) \tablenotemark{~b}} &
           \colhead{($L_{3000}$;~$\ell$) \tablenotemark{~b}} &
           \colhead{(\mbh;~$\ell$) \tablenotemark{~b}}
}
\startdata
\rs    \tablenotemark{~c} &   0.55     & $-$0.18  & 0.38     & $-$0.59  & $-$0.23  & $-$0.57  & $-$0.48  & 0.05  & $-$0.02 & 0.00 \\
\pnull \tablenotemark{~c} &   $<$1e-38 &  6e-16   & $<$1e-38 & $<$1e-38 & $<$1e-38 & $<$1e-38 & $<$1e-38 & 0.03  & 0.3     & 0.8  \\
\hline \\
\rs    \tablenotemark{~d} &   0.49     & $-$0.22  & 0.27     & $-$0.55  & $-$0.28  & $-$0.51  & $-$0.49  & 0.07  & 0.02    & 0.04 \\
\pnull \tablenotemark{~d} &   $<$1e-38 &  8e-25   & $<$1e-38 & $<$1e-38 & $<$1e-38 & $<$1e-38 & $<$1e-38 & 0.001 & 0.5     & 0.07
\enddata
\tablenotetext{a}{~For every (partial) correlation,
we list the Spearman rank correlation
coefficient (\rs) and the probability of the null hypothesis (\pnull)
for the 2092 objects in the sample.}
\tablenotetext{b}{~($Y$;\,$Z$) denotes the partial correlation
between the \mgii\ EW and $Y$, controlling for $Z$.}
\tablenotetext{c}{~Top panel: FWHM is measured from the model \mgii\ broad line;
$L$ is $\lambda L_{\lambda}$(3000\AA) as measured from the power-law model
of the local AGN continuum;
The black hole masses, \mbh, are calculated using
the formalism presented in McLure \& Dunlop (2004);
Eddington ratios ($\ell \equiv \lratio$) are calculated assuming
that the bolometric luminosity $\lbol \approx 5.9 \lambda L_{\lambda}$(3000\AA).}
\tablenotetext{d}{~Bottom panel: FWHM is measured from the model \hb\ broad line;
$L$ is $\lambda L_{\lambda}$(5100\AA) as measured from the power-law model
of the local AGN continuum;
The black hole masses, \mbh, are calculated using
the formalism presented in Collin et al. (2006, their Eqn. 7);
Eddington ratios are calculated assuming
that the bolometric luminosity $\lbol \approx 9\lambda L_{\lambda}$(5100\AA).}
\end{deluxetable}

\begin{deluxetable}{lccc}
\tablenum{2} \tablewidth{0pt} \topmargin 0.0cm \evensidemargin = 0mm
\oddsidemargin = 0mm
\tablecaption{
Fit Parameters of $\log L_{\rm \mgii}  ~=~  k \, \log L_{3000}   ~+~  b$
}
\tablehead{ \colhead{Sample} &
            \colhead{$N$}\tablenotemark{a} &
            \colhead{$k$} &
            \colhead{$b$}
}
\startdata
Full                     & 2092   & $0.91\pm0.01$ \tablenotemark{~b}         & $2.05\pm0.56$  \\
Full                     & 2092   & 1  \tablenotemark{~c}  & $-2.03\pm0.00$   \\
$\lratio \geq 0.8$       & 433    & $0.99\pm0.02$          & $-1.94\pm0.85$   \\
$\lratio \geq 0.8$       & 433    & 1  \tablenotemark{~c}  & $-2.19\pm0.00$   \\
$0.2 < \lratio <0.8$     & 1203   & $0.99\pm0.01$          & $-1.52\pm0.67$   \\
$0.2 < \lratio <0.8$     & 1203   & 1  \tablenotemark{~c}  & $-2.01\pm0.00$   \\
$\lratio \leq 0.2$       & 456    & 1  \tablenotemark{~c}  & $-1.83\pm0.00$   \\
$0.1 < \lratio \leq 0.2$ & 299    & $0.94\pm0.04$          & $0.55 \pm1.70$   \\
$0.1 < \lratio \leq 0.2$ & 299    & 1  \tablenotemark{~c}  & $-1.84\pm0.00$
\enddata
\tablenotetext{a}{~$N$ denotes the number of data points in every sample.}
\tablenotetext{b}{~For the full sample, $k$ is biased significantly
by the selection effect, see \S3.2.}
\tablenotetext{c}{~In these fittings $k$ is fixed to be 1.}
\end{deluxetable}

\end{document}